\documentclass[a4paper]{jpconf}
\usepackage{amsmath} 
\usepackage{amssymb} 
\usepackage{graphicx}
\usepackage{dcolumn}
\usepackage{bm}
\usepackage{xspace}
\usepackage{citesort}
\newcommand{\musr}{$\mu$SR\xspace}
\newcommand{\eup}{EuFe$_{2}$(As$_{1-x}$P$_{x}$)$_{2}$\xspace}
\newcommand{\mbs}{${}^{57}$Fe M\"ossbauer spectroscopy\xspace}

\newcommand{\mss}{${}^{57}$Fe\xspace}

\begin{document}
\title{Interplay of magnetism and superconductivity in EuFe$_{2}$(As$_{1-x}$P$_{x}$)$_{2}$ single crystals probed by muon spin rotation and ${}^{57}$Fe~M\"ossbauer spectroscopy} 

\author{T Goltz$^1$, S Kamusella$^1$, H S Jeevan$^2$, P Gegenwart$^{2,}$\footnotemark[-1], H Luetkens$^3$, P Materne$^1$, J Spehling$^1$, R Sarkar$^1$, H-H Klauss$^1$}
\address{$^1$Technical University of Dresden, Institute of Solid State Physics, Germany}
\address{$^2$I. Physik. Institut, Georg-August Universit\"at G\"ottingen, Germany}
\address{$^3$Laboratory for Muon Spin Spectroscopy, Paul Scherrer Institut, Villigen, Switzerland}
\ead{goltz@physik.tu-dresden.de}

\begin{abstract}
\footnotetext{\scriptsize Present adress: Exp. Physics VI, Center for Electronic Correlations and Magnetism, University of Augsburg, Germany}
We present our results of a local probe study on EuFe$_{2}$(As$_{1-x}$P$_{x}$)$_{2}$ single crystals with $x$=0.13, 0.19 and 0.28 by means of muon spin rotation and ${}^{57}$Fe M\"ossbauer spectroscopy. We focus our discussion on the sample with $x$=0.19 \emph{viz.} at the optimal substitution level, where bulk superconductivity ($T_{\text{SC}}=28$\,K) sets in above static europium order ($T^{\text{Eu}}=20$\,K) but well below the onset of the iron antiferromagnetic (AFM) transition ($\sim$100\,K). We find enhanced spin dynamics in the Fe sublattice closely above $T_{\text{SC}}$ and propose that these are related to enhanced Eu fluctuations due to the evident coupling of both sublattices observed in our experiments.
\end{abstract}

\section{Introduction}
The interplay of magnetism and superconductivity is one of the central topics in the contemporary studies on ferropnictides. Notably, the $A$Fe$_{2}$As$_{2}$-based compounds ($A$=Ba~\cite{RotterPhysRevLett.101.107006}, Sr~\cite{Schnelle-Sr122-masterpaper-PhysRevLett.101.207004} and Eu~\cite{Tegel-StrucUndMagEu122undSr122-NJP-0953-8984-20-45-452201}) have been widely studied since reasonably good single crystals can be obtained for chemical substitution on all three sites. 
Of particular interest within this so-called '122-family' is the superconducting \eup system for two reasons. Firstly, the substition of As by P is (nominally) isovalent thus superconductivity is not introduced by extra charge carriers and secondly, it contains a magnetic rare earth element on the $A$-site giving rise to magnetic order of the local Eu$^{2+}$ 4$f$ electrons in addition to the \emph{itinerant} antiferromagnetic iron order of the 3$d$ conduction band electrons.

For \eup, previous studies reported that the Fe AFM ordering and the accompanying structural transition from tetragonal to orthorhombic is suppressed upon P~substitution and eventually vanishes \emph{prior} to the appearance of a superconducting dome~\cite{JeevanPhaseDiagramPhysRevB.83.054511,ZapfPhysRevB.84.140503,CaoPhaseDiagram0953-8984-23-46-464204,Nowik-PdopedEu122-0953-8984-23-6-065701}.
In contrast, by measuring resistivity on a single crystal with $x$=0.13 under hydrostatical pressure, Tokiwa et al.~\cite{YoshiPhaseDiagramPhysRevB.86.220505} demonstrated the presence of a \emph{precursory} structural 
and Fe AFM transition above $T_{\text{SC}}$ between $p=0.4-0.8$\,GPa (refering to $x_{\text{P}}=0.15-0.20$). Likewise, a $\mu$SR pressure study on powdered samples by Guguchia et al.~\cite{GuguchiaMuSR-PdopedEu122} evidenced static magnetic order above the onset of superconductivity for similar pressures but they also conclude that the SDW ground state is differently affected by $x$ and $p$. Only recently, Nandi et al.~\cite{NandiPhysRevB.89.014512} showed the existence of a finite orthorhombic splitting reminiscent of weak Fe order~\cite{Goltz-PhysRevB.89.144511} below 50\,K in a superconducting ($T_{\text{SC}}=25$\,K) single crystal with $x_\text{P}=0.15$ at ambient conditions. 

Up to now, no comprehensive microscopic study of the ($T$-$x_\text{P}$) electronic phase diagram on \eup single crystals without any explicit symmetry-breaking forces (\emph{viz.} at zero external field and ambient pressure) is available to the best of our knowledge. Some microscopic studies on polycrystalline material were done by Nowik et al.~\cite{Nowik-PdopedEu122-0953-8984-23-6-065701} and Guguchia et al.~\cite{GuguchiaMuSR-PdopedEu122} but the reported transition temperatures from macroscopic measurements differ systematically comparing single crystalline \cite{JeevanPhaseDiagramPhysRevB.83.054511,ZapfPhysRevB.84.140503,ZapfSpinGlassPhysRevLett.110.237002} and polycrystalline \cite{CaoPhaseDiagram0953-8984-23-46-464204,Nowik-PdopedEu122-0953-8984-23-6-065701} samples. 
We only found one local probe study on a P-substituted \eup single crystal with $x$=0.3 ($T^{\text{mid}}_{\text{SC}}$=10.5\,K) by Munevar et al.~\cite{Munevar201418} but unfortunately they focussed on low temperatures and did not investigated temperatures above 50\,K.

In view of this gap this work emphazises, that further microscopic studies of single crystalline \eup in the full temperature range are needed to get a better understanding of the precursory (\textit{T}$>$\textit{T}$^{\text{Eu}}$) Fe order and its possible importance for the appearance of superconductivity.
Due to the lenght restriction of this article, we can only present the main results from the sample with $x$=0.19. A more detailed discussion of the obtained ($T$-$x_\text{P}$) phase diagram shown in Fig.\,\ref{fig:phasediagram} is in preparation~\cite{GoltzPhd}.
\section{Experimental Methods}
Single crystals of \eup were grown by the FeAs self-flux method. The homogeneity and actual composition of the three samples with $x$=0.13, $0.19$ and $0.28$ was confirmed within $\Delta x = 0.01$ error by EDX microprobe analysis on several points of the sample. Thermodynamic properties were determined by resistivity, magnetization and specific heat measurements according to~\cite{JeevanPhaseDiagramPhysRevB.83.054511} indicating bulk superconductivity for the sample with $x$=0.19 whereas the samples with $x$=0.13 and $0.28$ are non-superconducting.
$^{57}$Fe M\"ossbauer spectroscopy (MS) was performed in a standard transmission geometry setup using a $^{57}$Co/Rh source with an experimental line witdh~(HWHM) of $\omega=0.135(5)\,$mm/s. 
An Oxford LLD1 cryostat with a standard VTI was used to stabilize temperatures between 2\,K and 300\,K. M\"ossbauer spectra were evaluated by diagonalizing the full static hyperfine Hamiltonian including electric quadrupole and magnetic hyperfine interaction using the maximum entropy method (MEM) option to extract the iron hyperfine field distribution $\rho(\text{B})$ provided by the M\"ossFit package~\cite{BubelMoessfit}.
Muon spin rotation measurements were performed using the GPS spectrometer at the $\pi$M3 beamline of the Swiss Muon Source at the Paul Scherrer Institut, Switzerland. The data was analyzed with the \textsc{MUSRFIT} package~\cite{Suter2012}. In all \musr experiments, the plate-like single crystals were mounted with the crystal \emph{c}-axis parallel to the muon beam. 
The initial muon spin polarization was rotated by $-42^{\circ}$ with respect to the beam and the \emph{z}-direction of the laboratory framework as determined by transerse field (TF) experiments. In this arrangement, we can simultaneously measure the zero field (ZF) \musr time spectra for $\mu$ spin $\perp c$ and $\mu$ spin $\parallel c$ by evaluating the asymmetry signal from the up-down (UD) and forward-backward (FB) detector-pairs seperately. 
\section{Results}
%
\subsection{$x$=0.13 single crystal (batch no. HS03T2)}
Iron AFM ordering sets in below $115(5)$\,K and the magnetically ordered volume fraction gradually increases at lower temperatures and saturates at $\sim$65\% for T=40\,K as determined by weak~TF-$\mu$SR. We define the AFM transition temperature for this sample at the midpoint of the very broad transition  
leading to $T^{\text{Fe}}_\text{N}=71(3)$\,K. Consistently, \mss MS spectra show (i) a significant line broadening between 140 and 80\,K and (ii) a clear change in the hyperfine field distribution $\rho(\text{B})$ shifting about 75\% of the spectral weight to discernible hyperfine field values ($\rho(\text{B})>$1.25\,T) below 40\,K. This change in $\rho(\text{B})$ is reminiscent to the modification of the hyperfine field distribution W(B)\footnote{Note that when using the maximum entropy method, $\rho(\text{B})$ is deduced directly from the raw data whereas W(B) is constructed from the SDW shape, thus a priori includes an implicit premise about the underlying physics.} due to the alteration of the SDW shape from sinoidal to rectangular in EuFe$_{2}$As$_{2}$ between 191 and 145\,K reported by B\l{}achowski et al.~\cite{Blachowski-SDWshape-PhysRevB.83.134410}. Our MS data yield a mean hyperfine field value of $\left\langle B^{\text{Fe}}_\text{hf}\right\rangle=4.1(3)$\,T and a total spectral shift of IS=0.50(1)\,mm/s relative to metallic iron at $T$=4.2\,K. Independent of temperature, the local field is canted by $15(5)^{\circ}$ out of the ab-plane ($\theta^{\text{c-axis}}=75(5)^{\circ}$). 
%
%
%
%
\subsection{$x$=0.19 single crystal (batch no. HS03E)}
\begin{figure}[t]
\begin{center}
\includegraphics[width=0.32\textwidth]{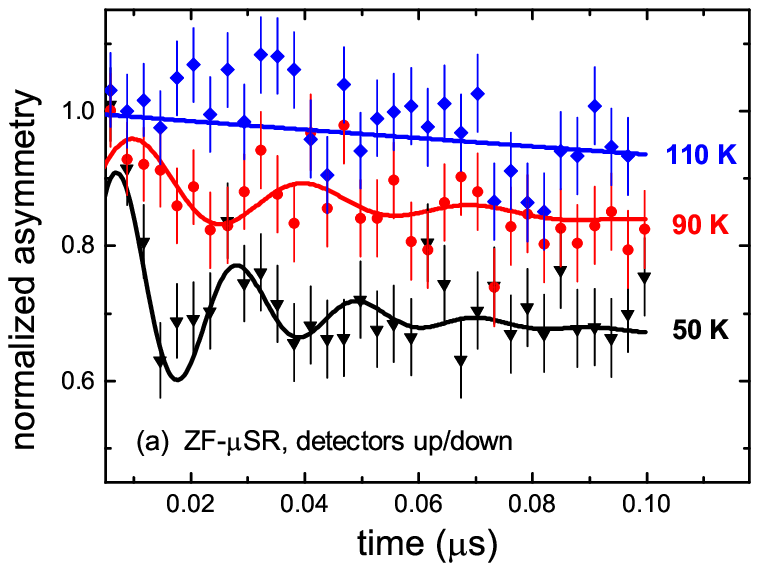}\hspace{0.5pc}		
\includegraphics[width=0.32\textwidth]{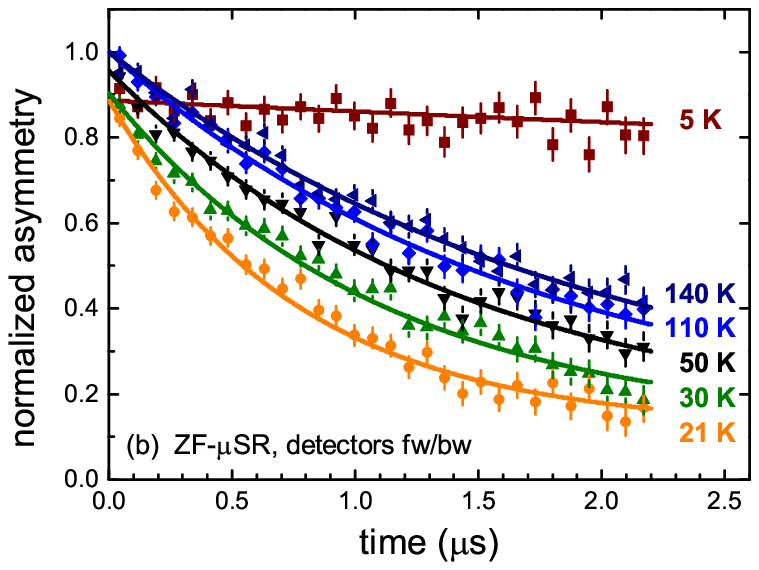}\hfill
\includegraphics[width=0.31\textwidth]{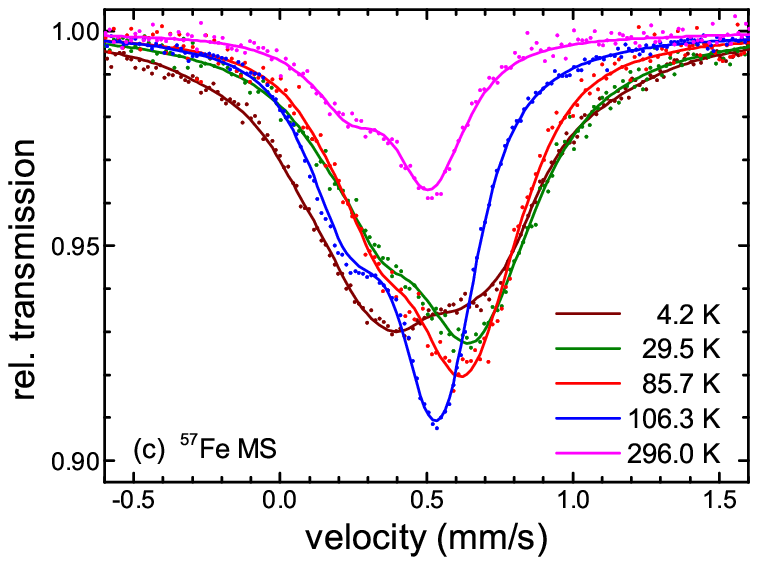}
\end{center}
\caption{\label{fig:x19zfmusr}Selective ZF-\musr and ${}^{57}$Fe M\"ossbauer spectra of \eup with $x$=0.19.}
\end{figure}

\begin{figure}[t]
\begin{minipage}{0.53\textwidth}
\includegraphics[width=19pc]{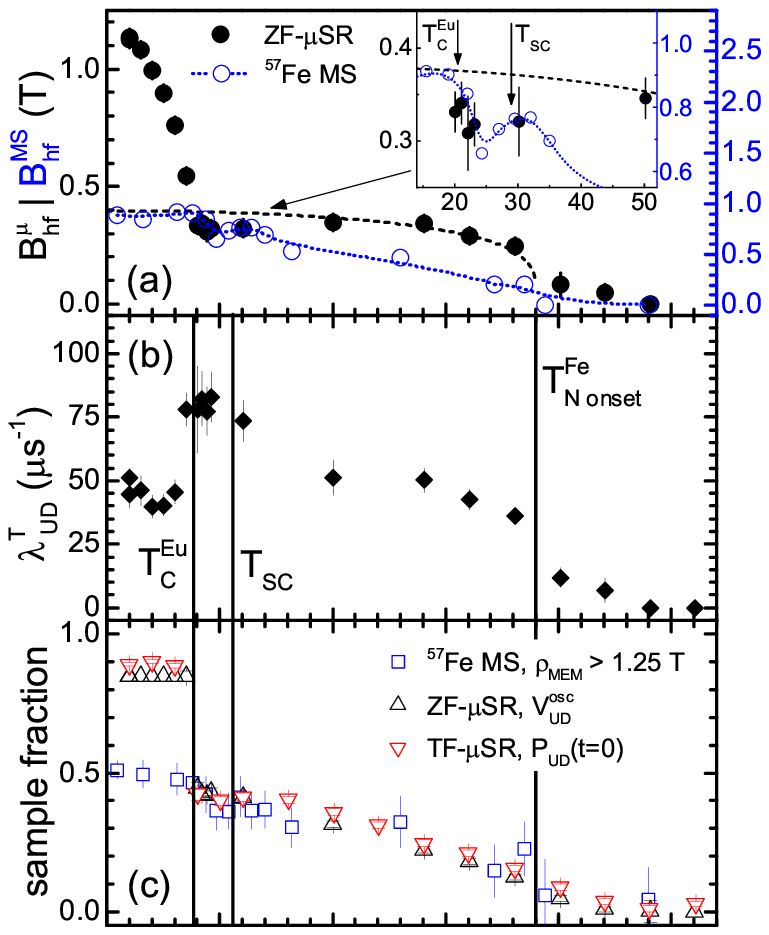}\\[0.1pc]
\includegraphics[width=18pc]{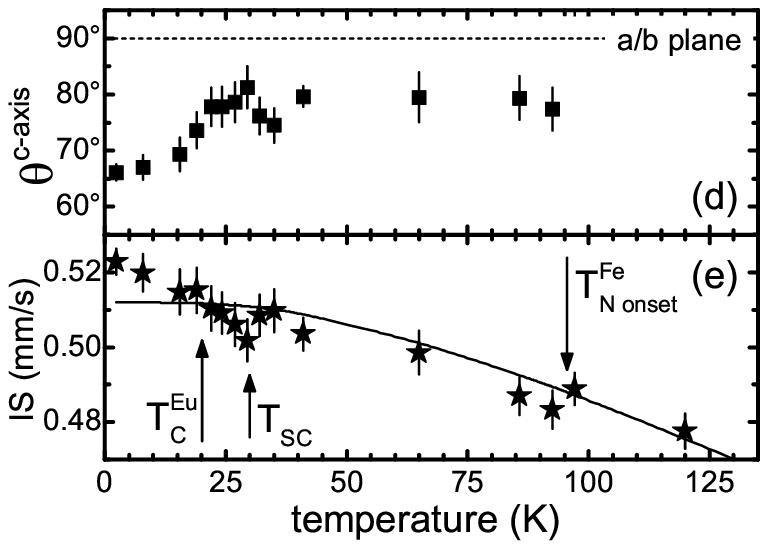}%
\end{minipage}	
\begin{minipage}{0.46\textwidth}\caption{\label{fig:x19overview}\\Compilation of the main results for \eup with $x$=0.19:\\[0.3pc]
(a) Local field at the muon site $B^\mu_{\text{hf}}$ (black dots, left scale) and on the $^{57}$Fe nucleus $B^{\text{MS}}_{\text{hf}}$ (blue dots, right scale). The black dashed line is a fit to a phenomenological order parameter function \cite{GuguchiaMuSR-PdopedEu122} given by
$B(T)=B_0 \cdot {\left[1 - {\left(T/T_{\text{N onset}}^{\text{Fe}}\right)}^\alpha  \right]}^\beta$
yielding $\alpha$=2, $\beta$=0.27(5) and $B_0$=0.4\,T. The dotted blue line is a guide to the eye, emphazising the small but reproducible decrease of $B^{\text{MS}}_\text{hf}$ below the superconducting transition temperature.\\[0.1pc]
(b) $\mu$SR transverse relaxation rate $\lambda_{\text{UD}}^T$\\[0.1pc]
(c) Sample fractions associated to the static Fe magnetic ordered volume fraction: ZF-$\mu$SR oscillatory signal $V_{\text{UD}}^{\text{osc}}$ (black triangles), total initial asymmetry P$_{\text{UD}}$(t=0) from weak~TF-$\mu$SR (red triangles) and integrated spectral weight with discernable hyperfine field values $\rho_{\text{MEM}}>1.25$\,T from ${}^{57}$Fe MS (blue squares).\\[0.1pc]
(d) Tilting angle $\theta^{\text{c-axis}}$ of the \mss MS hyperfine field 
with respect to the c-axis.\\[0.1pc]
(e) ${}^{57}$Fe MS total spectral shift (IS) relative to metallic iron. The line is a standart Debye-fit yielding 
$\theta_D$=260(110)\,K and 
M$_{\text{eff}}$=$68(11)$\,u. Note, that a slight decrease of IS below the onset of Fe order is also reported for single crystalline EuFe$_2$As$_2$~\cite{Blachowski-SDWshape-PhysRevB.83.134410}.
}
\end{minipage}
\end{figure}

ZF-\musr and ${}^{57}$Fe M\"ossbauer spectra are shown in Fig.\,\ref{fig:x19zfmusr}. The results of our analysis are compiled in Fig.\,\ref{fig:x19overview}. The ZF-\musr spectra at high temperatures (T$>$150\,K) show a very weakly damped exponential muon spin depolarization in both detector pairs. Below 100\,K, a strongly damped, cosine shaped precession signal becomes visible in the UD detector pair ($\mu$ spin $\perp c$) at short times marking the onset of static \emph{iron} magnetic ordering. 
The amplitude becomes more pronounced upon further cooling down to 30\,K showing a gradual increase of the magnetic volume fraction $V^{\text{osc}}$, see Fig.\,\ref{fig:x19overview}(c). Analogue to~\cite{GuguchiaMuSR-PdopedEu122} we used the muon spin depolarization function
\begin{equation}
	P_{\text{UD}}(t)= V_{\text{UD}}^{\text{osc}} \: \cos \left(2 \pi f_\mu \cdot t + \phi \right) e^{-\lambda_{\text{UD}}^T\cdot t } + \: V_{\text{UD}}^{\text{relax}} \: e^{-\lambda_{\text{UD}}^L\cdot t }
	\label{eq:x19zf} 
\end{equation}
to fit our data. The total initial asymmetry $P_{\text{UD}}(t=0)$ is normalized to $1$ at 150\,K. 
In the FB detector pairs ($\mu$ spin $\parallel c$), no oscillations can be observed for all temperatures. Instead, only a slow exponentially relaxing signal according to the second term in Eq.\,(\ref{eq:x19zf}) with a small decrease in the initial value of the normalized asymmetry $P_{\text{FB}}(t=0)$ is found. $P_{\text{FB}}(t=0)$ was likewise normalized to 1 at 150\,K and decreases continiously below 100\,K to 0.9 at 21\,K~($T>$\textit{T}$^{\text{Eu}}$), see Fig.\,\ref{fig:x19zfmusr}(b). 
From this follows, that the local field at the muon site due to static Fe ordering points essentially parallel~$c$ beeing suggestive of an ordered iron moment within the a/b plane. This finding is corroborated by our results from \mbs which yield a tilting angle of $\theta^{\text{c-axis}}=78(5)^{\circ}$ for the static iron hyperfine field with respect to the c-axis in the same temperature range as shown in Fig.\,\ref{fig:x19overview}(d).

Below 50\,K, we observe additional damping of the oscillatory signal in the ZF-$\mu$SR spectra. The transverse relaxation rate $\lambda_{\text{UD}}^T$ increases from 50 to 75\,$\mu s^{-1}$ closely above $T_{\text{SC}}$ and gets back to 50\,$\mu s^{-1}$ below $T^{\text{Eu}}$, see Fig.\,\ref{fig:x19overview}(b).

Below 20\,K, additive Eu FM ordering is evidenced by a much faster ($>$100\,MHz) precessing signal than for the iron order in the UD detector pair ($\mu$ spin $\perp c$) and a nearly constant muon spin depolarization in the FB detector pair ($\mu$ spin $\parallel c$). From the absolute asymmetry values, $V_{\text{UD}}^{\text{osc}}$ and P$_{\text{FB}}$(t=0) in the ordered (\textit{T}=1.6\,K) and normal state (\textit{T}$>$150\,K), we calculated~\cite{GoltzPhd} the c-axis tilting angle of the local field at the muon site to be $9(2)^{\circ}$ at \textit{T}=1.6\,K. 
The temperature dependence of $B^\mu_{\text{hf}}$ is shown in Fig.\,\ref{fig:x19overview}(a) together with $\left\langle B^{\text{MS}}_\text{hf}\right\rangle$ from \mss M\"ossbauer spectroscopy.
%
%
\subsection{$x$=0.28 single crystal (batch no. HS33/34/35)}
For \eup with $x$=0.28, we observe very weak iron magnetism below 120(5)\,K as deduced from significant line broadening of the \mss MS spectra using a pseudo-paramagnetic fit. More clearly, the ZF-$\mu$SR time spectra show weak Fe order seen by additional strong electronic relaxation ($\approx$50\,$\mu\text{s}^{-1}$) at short times for $T$=100\,K and below, which is not present at 150\,K. The relative intensity for this fast relaxation signal can be well separated and accounts for the magnetic volume fraction. It gradually increases from $\sim$4\% at 100\,K close to 20\% above the onset of static Eu ordering.

Below $T^{\text{Eu}}$=20\,K, ZF-$\mu$SR time spectra of \eup with $x$=0.28 are similar to those of the $x$=0.19 sample. The local field at the muon site lies essentially parallel to the c-axis and the Eu magnetism covers the complete sample volume. 
\mss MS spectra show an increasing transferred hyperfine field up to 1.2(1)\,T at 4.2\,K which is accompagnied by a continious spatial reorientation of the (total) hyperfine field. The tilting of $\left\langle B^{\text{MS}}_\text{hf}\right\rangle$ with respect to the c-axis changes from $\theta^{\text{c-axis}}$=$60^{\circ}$ at 19\,K to $40^{\circ}$ at \textit{T}=4.2\,K.
\clearpage
\section{Discussion}
\subsection{Precursory Fe order (\textit{T}$>$\textit{T}$_{\text{SC}}$, \textit{T}$>$\textit{T}$^{\text{Eu}}$) in \eup with $x$=0.13, 0.19 and 0.28}
\begin{figure}[t]
\begin{minipage}[b]{0.53\textwidth}
\includegraphics[width=1\textwidth]{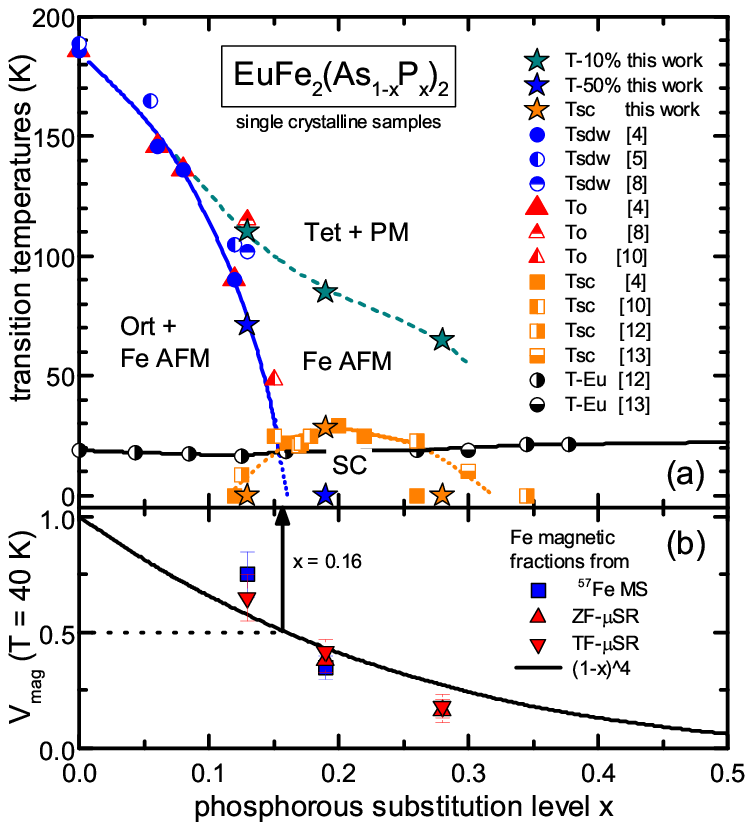}
\end{minipage}
\hfill
\begin{minipage}[b]{0.43\textwidth}
{
\caption{\label{fig:phasediagram}\\(a) ($T$-$x_\text{P}$)~phase diagram for single crystalline \eup under ambient conditions. Our finding of precursory Fe magnetism for $x$=0.19 and 0.28 is illustrated by $T_{\text{10\%}}$ (turquois). Structural (red triangles), Fe SDW (blue dots), static Eu (black dots) and superconducting (orange squares) phase transition temperatures from macroscopic methods are taken from the corresponding references. $T_{\text{10\%}}$ and $T_{\text{50\%}}$ are defined in the main article; Tet=tetragonal, Ort=orthorhombic.\\
(b)~Fe AFM magnetic volume fraction at \textit{T}=40\,K (\textit{T}$>$\textit{T}$^{\text{Eu}}$) in comparision to $P(x)$=${\left(1-x\right)}^4$ (see main article).\\
Bulk superconductivity emerges close to $x$=0.16 where $V_{\text{mag}}$ falls below the value of 50\% according to $(1-x)^4$ and consequently, $T_{\text{50\%}}$(=$T_{\text{SDW}}$) vanishes.}
}
\end{minipage}
\end{figure}

For $x$=0.13 and $x$=0.19, we find static iron magnetism below 
115(5) and 95(5)\,K. The Fe hyperfine field $\left\langle B^{\text{MS}}_\text{hf}\right\rangle$ as well as the magnetic volume fraction ($V_{\text{mag}}$) increases gradually upon cooling. At 40\,K (fairly above \textit{T}$^{\text{Eu}}$), we find $V_{\text{mag}}$=$70$\% and 40\% along with $B^{\text{MS}}_{\text{hf}}$=4.0 and 0.5\,T, respectively. For $x$=0.28, $V_{\text{mag}}$ does not exceed 20\% above $T^{\text{Eu}}$ and it remains unclear whether or not it is static on the $\mu$SR and M\"ossbauer timescale. 
In order to be sure not to overestimate the magnetic anomalies 
for the compilation of the ($T$-$x_\text{P}$)~phase diagram (Fig.~\ref{fig:phasediagram}), we define firstly $T_{\text{10\%}}$ as the temperature, where at least 10\% of our raw data signal displays magnetic behaviour. 
Secondly, since a thermodynamic phase transition temperature cannot be well-defined from the very gradual magnetic transition we find for our \eup samples, we relate $T_{\text{SDW}}$ from macroscopic measurements to $T_{\text{50\%}}$, defined by the temperature for which $V_{\text{mag}}$=50\%~(percolation threshold). 
In this context, we emphasize that for $x$=0.19 and $x$=0.28, $V_{\text{mag}}$ is always smaller than 50\%, so the observed weak iron magnetism might not be seen by resistivity or other macroscopic methods. 

The obtained values for $V_{\text{mag}}$ above \textit{T}$^{\text{Eu}}$ lie close to the probalistic expression $P(x)$=${\left(1-x\right)}^4$ which accounts for the number of Fe atoms surrounded by four As atoms, see Fig.~\ref{fig:phasediagram}(b). This connection was pointed out by Nowik et al.~\cite{Nowik-PdopedEu122-0953-8984-23-6-065701} who interpreted the two parts in terms of commensurate ($P(x)$) and incommensurate ($1-P(x)$) magnetic components assuming $V_{\text{mag}}$=100\% below a given temperature. 
Nevertheless, considering that $\mu$SR is a much more sensitive probe for small ($<0.1\,\mu_B$) magnetic moments, we can self-consistently substanciate our interpretation of the \mss MS spectra in terms of magnetic and non-magnetic sample fractions to our weak~TF-$\mu$SR data. Furthermore, the cosine shaped ZF-$\mu$SR precession signal we observe for $x$=0.19 typically accounts for commensurate magnetic order.
However, without contradiction to \cite{Nowik-PdopedEu122-0953-8984-23-6-065701}, we conclude that on an atomic lenght scale, \emph{coherent} and presumably static magnetism is related to iron atoms which are surrounded by As atoms only. 
\subsection{Superconductivity and magnetism in \eup with $x$=0.19 (\textit{T}$<$\textit{T}$_{\text{SC}}$, \textit{T}$>$\textit{T}$^{\text{Eu}}$)}
For $x$=0.19, we find a small but clearly discernible decrease of $\left\langle B^{\text{MS}}_\text{hf}\right\rangle$ and constant $V_{\text{mag}}$ below the superconducting transition temperature $T_{\text{SC}}$ as shown in Fig.\,\ref{fig:x19overview}(a) and~(c). This strongly suggests that magnetism and superconductivity compete for the same electrons at least in parts of the sample volume thus pointing to coexistence of superconductivity and Fe magnetism. The observed decrease in the M\"ossbauer hyperfine field is accompagnied by an increase of the field distribution width in $\rho_{\text{MEM}}$. 
Consistently, the ZF-$\mu$SR transverse relaxation rate $\lambda^T$ increases from 50 to 75\,$\mu s^{-1}$ above $T_{\text{SC}}$ (see Fig.\,\ref{fig:x19overview}(b)). Since $\lambda^T$ contains static and dynamic components, we conclude that close to $T_{\text{SC}}$ enhanced spin dynamics weaken the static character of the Fe order. 
A change in the dynamics of the Fe magnetic moment closely above $T_{\text{SC}}$ was reported by Munevar et al. for a superconducting $x$=0.3 single crystal~\cite{Munevar201418}.
However, for our $x$=0.19 sample we find that below \textit{T}$^{\text{Eu}}$, the value of $\lambda^T$ is steplike-wise restored to 50\,$\mu s^{-1}$, indicating that the Fe sublattice magnetization is (re-)stabilized by the static Eu~order. This leads us to the conclusion that the enhanced spin dynamics of the Fe sublattice below 40\,K should be rather referred to enhanced Eu fluctuations. An interplay of both sublattices is clearly evidenced by the increase of the ${}^{57}$Fe MS total spectral shift below \textit{T}$^{\text{Eu}}$ in~Fig.\,\ref{fig:x19overview}(e).\\[0.5pc]
\noindent To summarize, we investigated the ($T$-$x_\text{P}$)~electronic phase diagram for single crystalline \eup by a local probe study of three samples with $x$=0.13, 0.19 and 0.28 under ambient conditions. 
Fe magnetism was found for all samples and the magnetic volume fractions are found to be related to iron atoms which are surrounded by As atoms only. For $x$=0.19, our data suggests that magnetism coexists with superconductivity.
\ack
This work was financially supported by the German Research Foundation (DFG) within SPP 1458 (projects KL~1086/10-1 and GE~1640/4-2), GRK 1621 and grant no. SA~2426/1-1. Part of this work was performed at the Swiss Muon Source (Villigen, Switzerland). T.G. thanks Shibabrata Nandi, Shuai Jiang and Sina Zapf for valuable discussions.
\section*{References}
\bibliography{eu122iop}

\providecommand{\newblock}{}
\begin{thebibliography}{10}
\expandafter\ifx\csname url\endcsname\relax
  \def\url#1{{\tt #1}}\fi
\expandafter\ifx\csname urlprefix\endcsname\relax\def\urlprefix{URL }\fi
\providecommand{\eprint}[2][]{\url{#2}}

\bibitem{RotterPhysRevLett.101.107006}
Rotter M, Tegel M and Johrendt D 2008 {\em Phys. Rev. Lett.\/} {\bf 101}(10)
  107006

\bibitem{Schnelle-Sr122-masterpaper-PhysRevLett.101.207004}
Leithe-Jasper A, Schnelle W, Geibel C and Rosner H 2008 {\em Phys. Rev.
  Lett.\/} {\bf 101}(20) 207004

\bibitem{Tegel-StrucUndMagEu122undSr122-NJP-0953-8984-20-45-452201}
Tegel M, Rotter M, Weiss V, Schappacher F~M, P\"ottgen R and Johrendt D {\em
  Journal of Physics: Condensed Matter\/} {\bf 20} 452201

\bibitem{JeevanPhaseDiagramPhysRevB.83.054511}
Jeevan H~S, Kasinathan D, Rosner H and Gegenwart P 2011 {\em Phys. Rev. B\/}
  {\bf 83}(5) 054511

\bibitem{ZapfPhysRevB.84.140503}
Zapf S, Wu D, Bogani L, Jeevan H~S, Gegenwart P and Dressel M 2011 {\em Phys.
  Rev. B\/} {\bf 84}(14) 140503

\bibitem{CaoPhaseDiagram0953-8984-23-46-464204}
Cao G, Xu S, Ren Z, Jiang S, Feng C and Xu Z~A {\em Journal of Physics:
  Condensed Matter\/} {\bf 23} 464204

\bibitem{Nowik-PdopedEu122-0953-8984-23-6-065701}
Nowik I, Felner I, Ren Z, Cao G~H and Xu Z~A {\em Journal of Physics: Condensed
  Matter\/} {\bf 23} 065701

\bibitem{YoshiPhaseDiagramPhysRevB.86.220505}
Tokiwa Y, {H\"ubner} S~H, Beck O, Jeevan H~S and Gegenwart P 2012 {\em Phys.
  Rev. B\/} {\bf 86}(22) 220505

\bibitem{GuguchiaMuSR-PdopedEu122}
Guguchia Z, Shengelaya A, Maisuradze A, Howald L, Bukowski Z, Chikovani M,
  Luetkens H, Katrych S, Karpinski J and Keller H 2013 {\em Journal of
  Superconductivity and Novel Magnetism\/} {\bf 26} 285--295

\bibitem{NandiPhysRevB.89.014512}
Nandi S, Jin W~T, Xiao Y, Su Y, Price S, Shukla D~K, Strempfer J, Jeevan H~S,
  Gegenwart P and Br\"uckel T 2014 {\em Phys. Rev. B\/} {\bf 89}(1) 014512

\bibitem{Goltz-PhysRevB.89.144511}
Goltz T, Zinth V, Johrendt D, Rosner H, Pascua G, Luetkens H, Materne P and
  Klauss H~H 2014 {\em Phys. Rev. B\/} {\bf 89}(14) 144511

\bibitem{ZapfSpinGlassPhysRevLett.110.237002}
Zapf S, Jeevan H~S, Ivek T, Pfister F, Klingert F, Jiang S, Wu D, Gegenwart P,
  Kremer R~K and Dressel M 2013 {\em Phys. Rev. Lett.\/} {\bf 110}(23) 237002

\bibitem{Munevar201418}
Munevar J, Micklitz H, Alzamora M, Argüello C, Goko T, Ning F, Munsie T,
  Williams T, Aczel A, Luke G, Chen G, Yu W, Uemura Y and Baggio-Saitovitch E
  2014 {\em Solid State Communications\/} {\bf 187} 18 -- 22

\bibitem{GoltzPhd}
Goltz T {Ph.D.~Thesis, TU Dresden} {(in preparation)}

\bibitem{BubelMoessfit}
Kamusella S {M\"ossFit: A Free Framework for {$^{57}$Fe} M\"ossbauer Data
  Analysis} {(private communication)}

\bibitem{Suter2012}
Suter A and Wojek B 2012 {\em Physics Procedia\/} {\bf 30} 69 -- 73 (12th
  {$\mu$SR} Conference, 2011)

\bibitem{Blachowski-SDWshape-PhysRevB.83.134410}
B\l{}achowski A, Ruebenbauer K, \ifmmode~\dot{Z}\else \.{Z}\fi{}ukrowski J,
  Rogacki K, Bukowski Z and Karpinski J 2011 {\em Phys. Rev. B\/} {\bf 83}(13)
  134410

\end{thebibliography}
\end{document}